# The Accuracy of Confidence Intervals for Field Normalised Indicators[1]

Mike Thelwall, Ruth Fairclough
Statistical Cybermetrics Research Group, University of Wolverhampton, UK.

When comparing the average citation impact of research groups, universities and countries, field normalisation reduces the influence of discipline and time. Confidence intervals for these indicators can help with attempts to infer whether differences between sets of publications are due to chance factors. Although both bootstrapping and formulae have been proposed for these, their accuracy is unknown. In response, this article uses simulated data to systematically compare the accuracy of confidence limits in the simplest possible case, a single field and year. The results suggest that the MNLCS (Mean Normalised Log-transformed Citation Score) confidence interval formula is conservative for large groups but almost always safe, whereas bootstrap MNLCS confidence intervals tend to be accurate but can be unsafe for smaller world or group sample sizes. In contrast, bootstrap MNCS (Mean Normalised Citation Score) confidence intervals can be very unsafe, although their accuracy increases with sample sizes.
**Keywords**: Citation analysis; field normalised citation indicators; confidence intervals

## 1 Introduction

Citation indicators that estimate the average citation rate of articles produced by a group are widely used in research assessment and for ranking universities, countries and departments (Aksnes, Schneider, & Gunnarsson, 2012; Albarrán, Perianes-Rodríguez, & Ruiz-Castillo, 2015; Braun, Glänzel, & Grupp, 1995; Elsevier, 2013; Fairclough & Thelwall, 2015). For example, in the U.K., they have been proposed for the national Research Excellence Framework (REF) to cross-check peer review judgements (Stern, 2016). If average citation indicators are to be used in such a role, then they must be calculated in a fair way and accompanied with an estimate of statistical variability so that strong conclusions are not drawn from small or biased differences.

Field normalised citation impact indicators adjust average citation counts for the field and year of publication to allow fair comparisons of citation impact between sets of articles that were published in different combinations of fields and years. For example, if group A published 100 medical humanities articles in 2014 with an average of 4 citations each but group B published 100 oncology articles in 2013 with an average of 30 citations each then it is not clear which had generated the most impactful research. Group B has two advantages: its articles are older, with longer to attract citations, and it publishes in an area where citations accrue rapidly. A field normalised indicator may divide by the average number of citations for the field and year so that the normalised counts are 1 if the average citation impact is equal to the world average. After this, it would be reasonable to compare the field normalised values of A and B. Nevertheless, confidence intervals or statistical hypothesis tests are needed to be able to judge whether the difference between A and B is likely to reflect an underlying trend rather than a random fluctuation of the data.

The use of statistical inference or confidence intervals to compare the average citation impact is uncommon within scientometrics and there are arguments against it, such as a lack of clarity about what exactly is being sampled (Waltman, 2016). Statistical inference is typically used when data is available about a sample whereas in scientometrics, relatively complete sets of publications are normally analysed and so there is no necessity to infer population properties from a sample, at least in the obvious sense. Nevertheless, research is a social process and therefore each citation is the product of activities that are affected by processes that can be thought of as random in the sense of not predictable in advance (Williams & Bornmann, 2016). The exact citation count of an article is therefore partly a result of chance factors rather than just the quality or value of an article. For example, if two essentially identical papers are published at the same time then one may become more highly cited than the other for spurious reasons, such as the prestige of the publishing journal (Larivière & Gingras, 2010), or the extent to which the citing literature is covered by the database used for the counts (Harzing & Alakangas, 2016; Table 3 in: Kousha & Thelwall, 2008). Thus, it seems impossible to regard citation counting as precisely measuring the impact of publications and it seems better to regard it instead as an inaccurate estimate (see the similar argument in: Waltman Traag, 2017). Moreover, the purpose of research evaluation is often to make decisions about future funding allocations or strategies based on past performance. In this context, the exact citation count of a paper is less important than the underlying capacity of a group to produce impactful research. Each article produced by a group can also be thought of as the product of both the underlying research power of the group and chance factors that affect the value of each paper produced. These chance factors include creativity-related factors that are internal to the researchers (Simonton, 2004) as well as external factors that are partly outside of their control, such as whether external technical or social developments turn their topic into one of societal importance (e.g., the recent rise in the importance of Arabic natural language processing and Middle Eastern studies). Thus, for example, Nobel Prize winners may occasionally produce rarely-cited research even if most of their output has high impact. In both contexts, statistical inference is reasonable and aligns with the standard social sciences practice of treating the situation as having an apparent population of plausible outcomes from the known parameters (Berk, Western, & Weiss, 1995; Bollen, 1995).

There are two alternative reasonable strategies to generate confidence limits. The parametric strategy assumes that the data follows a specific statistical distribution and then derives confidence limit formulae from an analysis of this distribution. The bootstrapping strategy resamples from the existing data, with replacement, and then calculates confidence limits in order that 95% (say) of the resampled indicator values fall within them (Efron & Tibshirani, 1986). Neither approach is perfect. The parametric strategy is reliant upon the distribution assumption and may also involve additional assumptions, such as that the distribution of a discretised distribution is like the continuous distribution that it was derived from. Bootstrapping is also unreliable for many data distributions and tasks (Hall, 1992; Hillis & Bull, 1993) and seems to be particularly unsuited to highly skewed data sets, such as those based on untransformed citation counts. In this context, it is not clear whether bootstrapping or parametric formulae are preferable for any given indicator and whether the optimal choice depends on basic properties of the data.

This article assesses the accuracy of bootstrapping for the calculation of confidence intervals for two field normalised average citation indicators. The Mean Normalised Citation Score (MNCS) (Waltman, van Eck, van Leeuwen, Visser, & van Raan, 2011ab), is used in the Leiden university ranking (Waltman, Calero-Medina, Kosten, Noyons, Tijssen, et al., 2012), and the Mean Normalised Log-transformed Citation Score (MNLCS) (Thelwall, 2017) is a more recent variant. This study focuses on a single field and year for pragmatic reasons: to allow an exploration of the impact of the mean and standard deviation without generating unmanageably many results from experiments with multiple fields and/or years. Confidence interval formulae have been proposed for the MNLCS and so these are also assessed for accuracy at the same time. Although there are many other field normalised indicators, these represent two of the main variants, with MNCS being well known and MNLCS being designed as a logical extension to deal with skewing in citation count data. One recent quite different indicator is the Relative Citation Ratio (RCR) (Hutchins, Yuan, Anderson, & Santangelo, 2016) but this is not included because it is not clear that it is relevant outside of biomedical science and its design makes bootstrapping highly complex because a paper's citations and the impact factors of the publishing journals for their references need to be modelled.

## 2  Background

The parametric strategy in statistics requires an assumption about the distribution of a citation data set. It has been known for a long time that citation counts diverge substantially from the normal distribution (de Solla Price, 1965) and that the power law is a much better fit if articles with few citations are ignored (Clauset, Shalizi, & Newman, 2009). Since field normalised indicators do not omit rarely cited articles and these often form the clear majority within a collection, the power law is an inappropriate distribution (Thelwall & Wilson, 2014a). Instead, both the discretised lognormal distribution and the hooked power law are reasonable fits for most sets of articles from a single field (or large monodisciplinary journal) and year (Radicchi & Castellano, 2012; Thelwall, 2016ab). Many alternative distributions and approaches have also been tested on the full range of citation counts, but none are clearly better than the discretised lognormal or hooked power law and most are worse, when fully tested. Appropriate stopped sum models have been found to fit citation data reasonably, but there is limited evidence of this and their parameters are too unstable to be useful in practice (Low, Wilson, & Thelwall, 2016). Negative binomial regression has also been used for citation data (Harhoff, Narin, Scherer, & Vopel, 1999) but the negative binomial distribution fits less well (Thelwall & Wilson, 2014b), including zero inflated variants (Low, Wilson, & Thelwall, 2016). The zero inflated Poisson distribution also does not fit well (Low, Wilson, & Thelwall, 2016).

In practice, there is little to choose between the discretised lognormal and hooked power law distributions and they have broadly similar shapes. Thus, it is reasonable to choose either as the basis for simulating citation data. Here, the discretised lognormal will be used because it is easier to manipulate its parameters independently. Parameter manipulation is also more easily interpreted for the discretised lognormal because its parameters are approximately the mean and standard deviation of the distribution after a logarithmic transformation.

### *2.1  The lognormal distribution*

The probability density function (PDF) for the continuous lognormal distribution (Limpert, Stahel, & Abbt, 2001) is as follows, where $\mu$ is a location parameter and $\sigma$ is a scale parameter.

$$\frac{1}{x\sigma\sqrt{2\pi}} e^{-\frac{(\ln(x)-\mu)^2}{2\sigma^2}} \quad (1)$$

The continuous lognormal distribution can be converted into a discrete distribution to match citation count data in two ways. First, the pdf of the continuous lognormal distribution can be treated as a probability mass function (PMF), after dividing by the sum of all PMF values. This adjustment is necessary for the PMF to sum to 1.

$$\frac{1}{x\sigma\sqrt{2\pi}} e^{-\frac{(\ln(x)-\mu)^2}{2\sigma^2}} \bigg/ \sum_{t=1}^{\infty} \frac{1}{t\sigma\sqrt{2\pi}} e^{-\frac{(\ln(t)-\mu)^2}{2\sigma^2}} \quad (2)$$

There is a problem with citation counts of zero, which cannot be modelled by a PMF designed in this way, because the PDF is undefined at zero. The standard solution to this, and the continuous variant below, is to add 1 to all citation counts so that zeros are avoided. The second method to convert the PDF into a PMF is to integrate the unit interval around each integer as follows, where the overall denominator again ensures that the sum of the PMF is 1.

$$\int_{x-0.5}^{x+0.5} \frac{1}{t\sigma\sqrt{2\pi}} e^{-\frac{(\ln(t)-\mu)^2}{2\sigma^2}} dt \bigg/ \int_{0.5}^{\infty} \frac{1}{t\sigma\sqrt{2\pi}} e^{-\frac{(\ln(t)-\mu)^2}{2\sigma^2}} dt \quad (3)$$

There is no evidence about which approach tends to fit empirical distributions better and in any case the choice makes little difference, especially for high citation counts, and especially for simulation exercises that vary the free parameters. The former (2) was therefore used as the main distribution here and the latter (3) to check that this choice did not affect the findings.

## 2.2 The Mean Normalised Log-transformed Citation Score

Mean Normalised Citation Score (MNCS) is the arithmetic mean of $\{c_i/l_i\}$ where $c_i$ is the citation count of paper $i$ and $l_i$ is the mean citation count of all papers published in the same year and field as $i$. Assuming the citation counts $c_i$ approximately follow the discretised lognormal distribution (2 or 3), which does not have a finite mean for some parameter values, there is no general formula for a confidence interval for the mean of the $c_i/l_i$ values.

The Mean Normalised Log-transformed Citation Score (MNLCS) is the arithmetic mean of $\left\{\frac{1}{L_i}\ln(1+c_i)\right\}$, where $L_i$ is the arithmetic mean of the $ln(1+c)$ log transformed citation counts for all articles in the same field and year as article $i$.

## 2.3 MNLCS Confidence Intervals

If the citation counts follow the discretised lognormal distribution, the log-transformed values may be close to the normal distribution and so a confidence interval formula has been proposed for the MNLCS based on this assumption (versions for the continuous lognormal in a non-ratio context are not appropriate, e.g., Zhou & Gao, 1997). The first formula assumes that the world average citation count is exact whereas the group papers form a sample, with sample deviation $s$ (Thelwall, 2017).

$$\text{MNLCS}_L = \text{MNLCS} - t_{n-1,\alpha} s/\sqrt{n} \quad (4a)$$
$$\text{MNLCS}_U = \text{MNLCS} + t_{n-1,\alpha} s/\sqrt{n} \quad (4b)$$

The second formula treats both the group's papers and the world's papers as samples so that both numerator and denominator are variable. The confidence interval is defined only if $h < 1$ in the formula. The quantities $SE_g$ and $SE_w$ are the standard errors and $\overline{c_g}$ and $\overline{c_w}$ are the arithmetic means of the group and world log-transformed citations $\ln(1+c)$, respectively.

$$h = \left(t_{n_1+n_2-2,\alpha} \frac{SE_w}{\overline{c_w}}\right)^2 \tag{5a}$$

$$SE_{\text{MNLCS}} = \frac{\overline{\text{MNLCS}}}{1-h}\sqrt{(1-h)\frac{SE_g^2}{\overline{c_g}^2} + \frac{SE_w^2}{\overline{c_w}^2}} \tag{5b}$$

$$\text{MNLCS}_{FL} = \frac{\overline{\text{MNLCS}}}{1-h} - t_{n_1+n_2-2,\alpha}SE_{\text{MNLCS}} \tag{5c}$$

$$\text{MNLCS}_{FU} = \frac{\overline{\text{MNLCS}}}{1-h} + t_{n_1+n_2-2,\alpha}SE_{\text{MNLCS}} \tag{5d}$$

Both formulae assume that the discretisation process in moving from the continuous to the discretised lognormal distribution does not affect the result much and the same for adding 1 before the log transformation.

# 3 Methods

To assess the accuracy of the MNLCS confidence interval formula and bootstrap confidence intervals for MNCS and MNLCS, the overall research design was to simulate many citation datasets using different discretised lognormal distribution parameters (see section 2 for a justification of the use of this distribution) and to derive confidence intervals for the mean of a specified subgroup for each sample. These were calculated from (a) the formula and (b) bootstrapping, comparing both to the estimated exact confidence intervals from 10,000 simulations with the same parameter set. For bootstrapping, confidence limits were calculated based upon 1000 resamples of the data set. This approach is consistent with the theoretical assumption of random sampling at the level of articles rather than individual citations (see: Waltman, 2016).

## 3.1 Location and scale parameters

The discretised lognormal location and scale parameters were varied to incorporate a range of values for recent and old articles, and for high and low citation fields, drawing on the range of parameters found for sets of articles from a single field and year in previous studies of articles that are about a decade old (Thelwall, 2016ab). These studies found parameter values in the range $0.5 < \mu < 5$ and $0.75 < \sigma < 1.5$. Since no studies seem to have fitted the discretised lognormal distribution to younger articles, this was done for sets of article with three years to attract citations (see Appendix, Table A1). Except for one outlier, these sets have much lower μ parameters and return values in the range $-0.8 < \mu < 1.8$ and $0.75 < \sigma < 1.5$. These overlap and so a single range of parameters was chosen to encompass both and extend the μ parameter to lower values: $-2 < \mu < 5$ and $0.75 < \sigma < 1.5$. These parameters cover a wide range of distributions, including very young articles. For example, at the lowest end of the spectrum, the first simulation for $\mu = -2, \sigma = 0.75$ and a world sample size of 5000 gave 4983 zeros and 17 ones, an average of 0.0034 citations per paper. At the other extreme, the first simulation with $\mu = 5, \sigma = 1.5$ gave citation counts from 0 (before adding 1) to 46548, with an arithmetic mean of 469 citations per paper.

## 3.2 Sample Size and Potential Sources of Differences

In addition to the distribution parameters, the size of the world set (i.e., all articles in a field and year, irrespective of their authorship) and group set can vary, as can the relationship between the average citation count of the group set and world set. To simplify the analysis, the subject sample size was fixed at 5000 for the world set and 500 for the subgroup for three of the five parameter sets (i.e., a 10% subgroup). This broadly corresponds to a medium sized field category in Scopus and a large subgroup, such as the articles of a major country. To set this in context, 25% of the 2,504,200 Web of Science core collection articles published in 2016 had an author from the USA, followed by China: 16%; the UK: 8%; Germany: 6%; Japan, India, France, Italy, Canada, Australia: 4%; Spain, South Korea, Brazil: 3%; Russia, The Netherlands, Turkey, Iran, Switzerland, Poland, Sweden: 2%. To analyse possible sources of differences, the following tests were run.

- **Basic set**: World parameter values corresponding to all values likely to be found in sets of articles from the same field and year (see Section 3.1), and at least a year old: $-2 < \mu < 5$ and $0.75 < \sigma < 1.5$; group parameter values the same; world sample size 5000; group sample size 500. Discretised lognormal distribution used to sample.
- **Differing group citation impact**: As for the basic set but with a group location parameter µ that is 0.5 larger than the world location parameter.
- **Smaller group sample size**: As for the basic set but with a group sample size of 250 (i.e., 5%, roughly corresponding to Germany, Japan, India, France, Italy, Canada, Australia).
- **Continuous distribution PMF**. As for the basic set but using the continuous lognormal distribution as a PMF.
- **Different world sample sizes**. As for the basic set but with $\mu = 1$ and the world sample size varying from 500 (a small subject category) to 10000 (a large subject category).

The expected value of each indicator was calculated by running the simulation 10000 times and taking the average (arithmetic mean) value of the group MNLCS and the same for the MNCS. This approach was used rather than calculating an exact expected MNLCS and MNCS value from the distribution parameters in case rounding errors in the R software used (which can be a problem in this context: Thelwall, 2016b) or other arithmetical problems could produce misleading theoretical values. Confidence intervals for the MNLCS and MNCS were calculated using bootstrapping (resampling 1000 times) for each of the 10000 iterations (10,000,000 MNLCS and MNCS calculations for each parameter set). The MNLCS formula (5) was also used to calculate 95% confidence intervals for each of the 10000 iterations. After all iterations had been completed, the percentage of confidence intervals containing the correct MNLCS or MNCS value (i.e., the average over 10000 iterations for the parameter set) was calculated. If the confidence intervals are precise then this figure should be close to 95% in all cases.

The R code used is available online (The *indicators for confidence intervals* R file at https://github.com/MikeThelwall/Informetrics_R_Code) and additional data and graphs for parameter values not illustrated below are available at: https://figshare.com/s/2378cdc027a8f8c303f3.

## 4  Results

For the basic set, the MNLCS formula confidence intervals contain the correct value more than 95% of the time overall, without substantial deviations for any parameter set (Figures 1 and 2). Thus the confidence interval is conservative, but reliable. In contrast, the bootstrapping confidence intervals are approximately correct for all parameter values for MNLCS. The MNLCS formula conservatism is due to the conservative assumption that the world population is independent of the group sample, whereas the group sample is a 10% subsample of the world population in this set.

The MNCS bootstrap confidence intervals are accurate for low values of $\sigma$ (Figure 1) but optimistic and unsafe for higher $\sigma$ values (Figure 2).

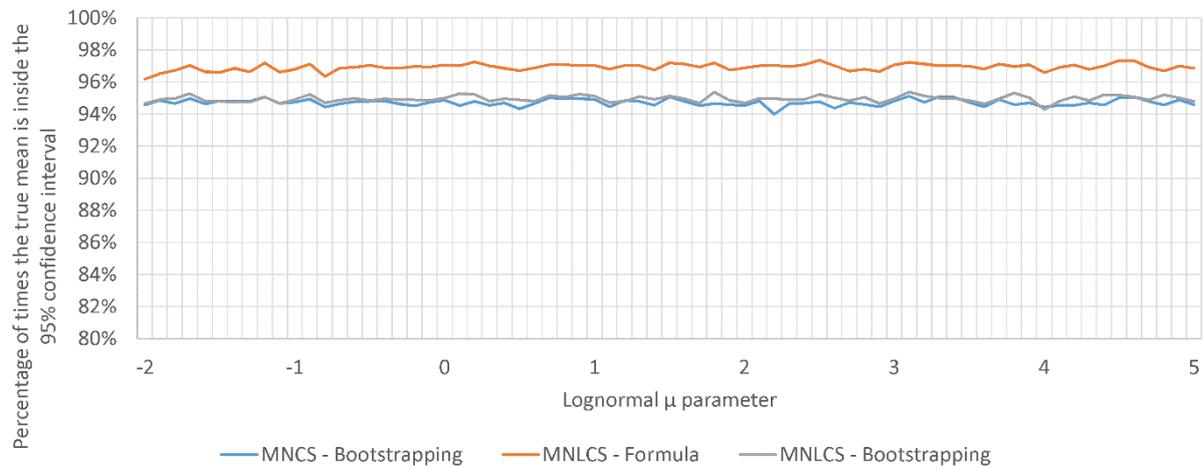

**Figure 1**. The proportion of the 95% confidence intervals calculated that contain the correct MNLCS or MNCS values based upon 10000 simulations at each set of parameter values (basic set with $\sigma = 0.75$, roughly corresponding to the UK or China publishing in a medium-sized subject category with low citation variability, for which their research has the same impact as the world average).

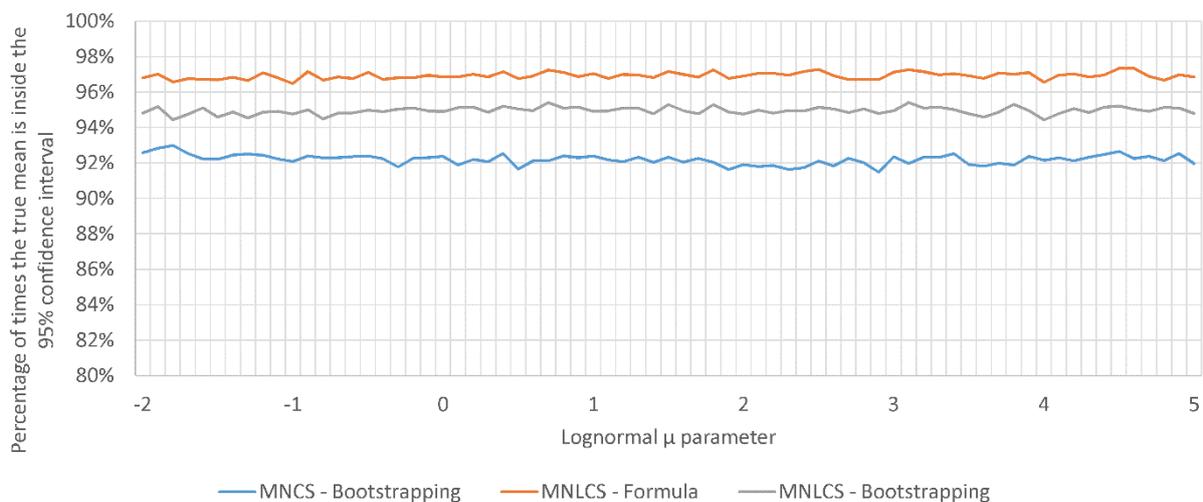

**Figure 2**. As for Figure 1 except that $\sigma = 1.5$, roughly corresponding to the UK or China publishing in a medium-sized subject category with *high* citation variability, for which their research has the same impact as the world average.

When the group mean is higher than the world mean (Figures 3, 4), the bootstrap confidence intervals are not affected but the formula confidence intervals become more conservative because the group sample has more influence on the world set due to its higher values.

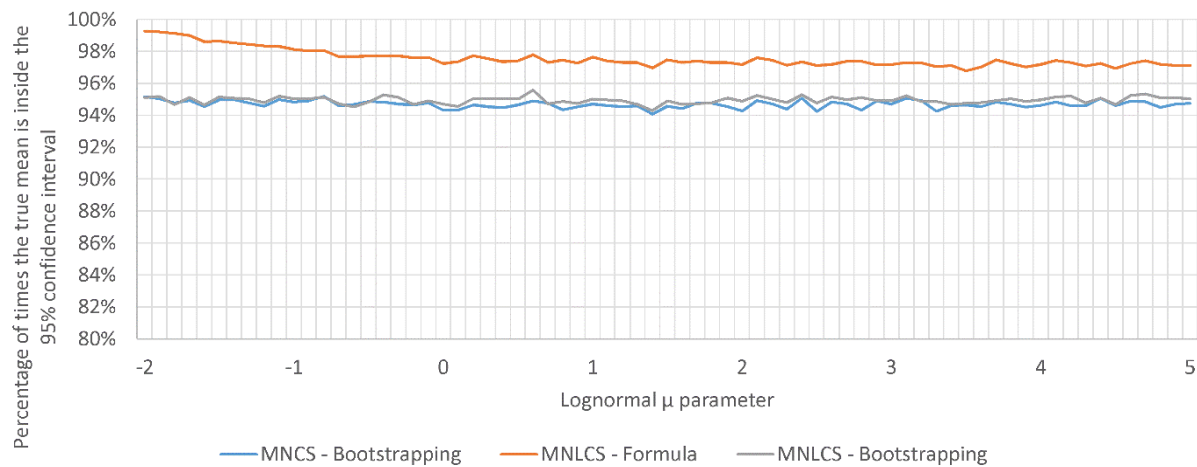

**Figure 3**. As for Figure 1 except for the differing group citation impact set, roughly corresponding to the UK or China publishing in a medium-sized subject category with low citation variability, for which their research has impact substantially *above* the world average.

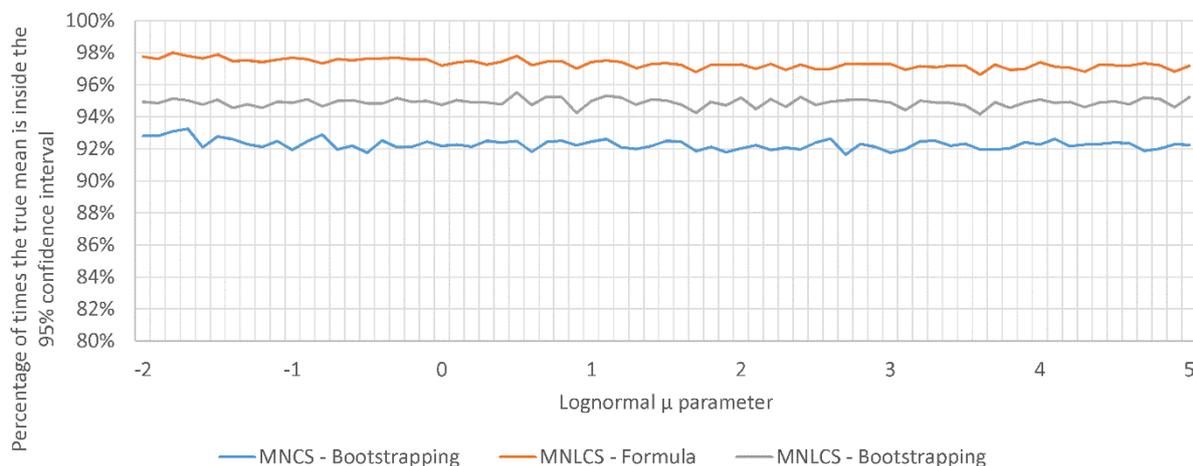

**Figure 4**. As for Figure 3 except $\sigma = 1.5$, roughly corresponding to the UK or China publishing in a medium-sized subject category with *high* citation variability, for which their research has impact substantially *above* the world average

When the group sample size is smaller relative to the world sample size (Figures 5, 6) the main difference is that the MNLCS formula confidence interval becomes less conservative. This confirms that its conservatism is due to the relative size of the group set within the world set.

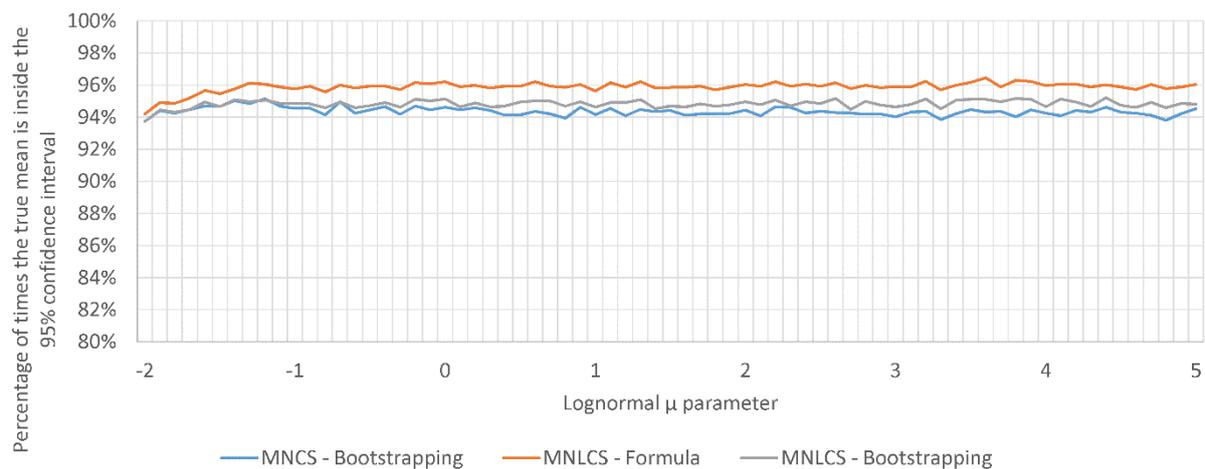

**Figure 5**. As for Figure 1 except for the smaller group sample size set (n=250 for the group), roughly corresponding to Germany, Japan, India, France, Italy, Canada, or Australia publishing in a medium-sized subject category with low citation variability, for which their research has the same impact as the world average.

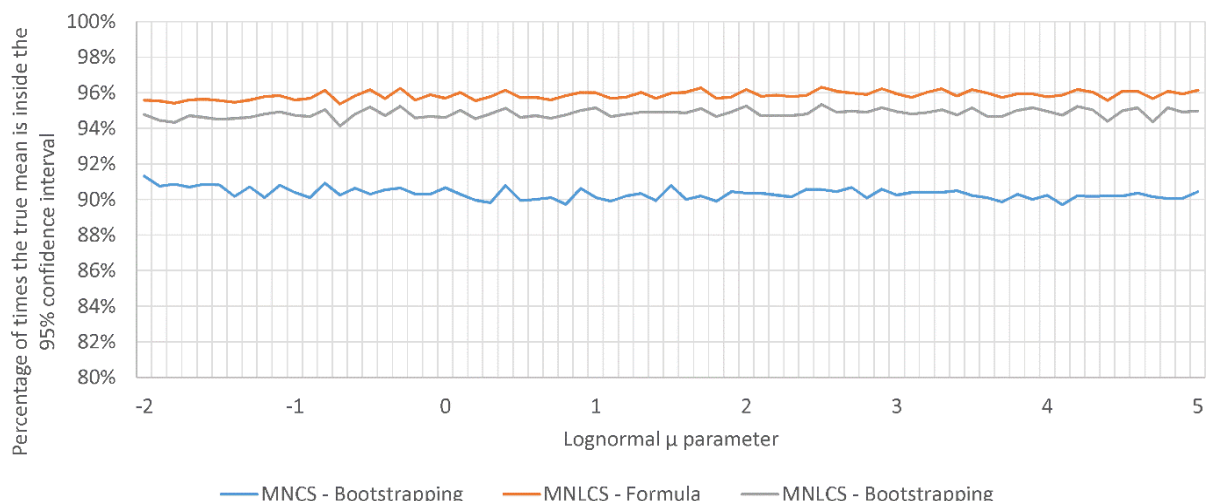

**Figure 6**. As for Figure 5 except $\sigma = 1.5$, roughly corresponding to Germany, Japan, India, France, Italy, Canada, or Australia publishing in a medium-sized subject category with *high* citation variability, for which their research has the same impact as the world average.

Switching from the discretised lognormal to the continuous lognormal treated as a PDF (Figures 7, 8) makes no difference to the results so they are not dependent upon the specific formula used.

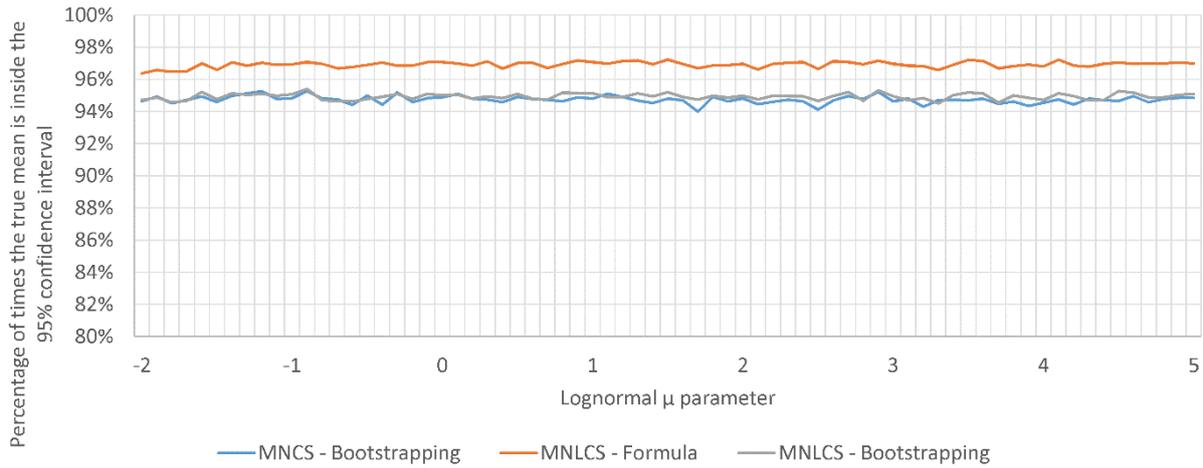

**Figure 7**. As for Figure 1 except for the continuous distribution PMF.

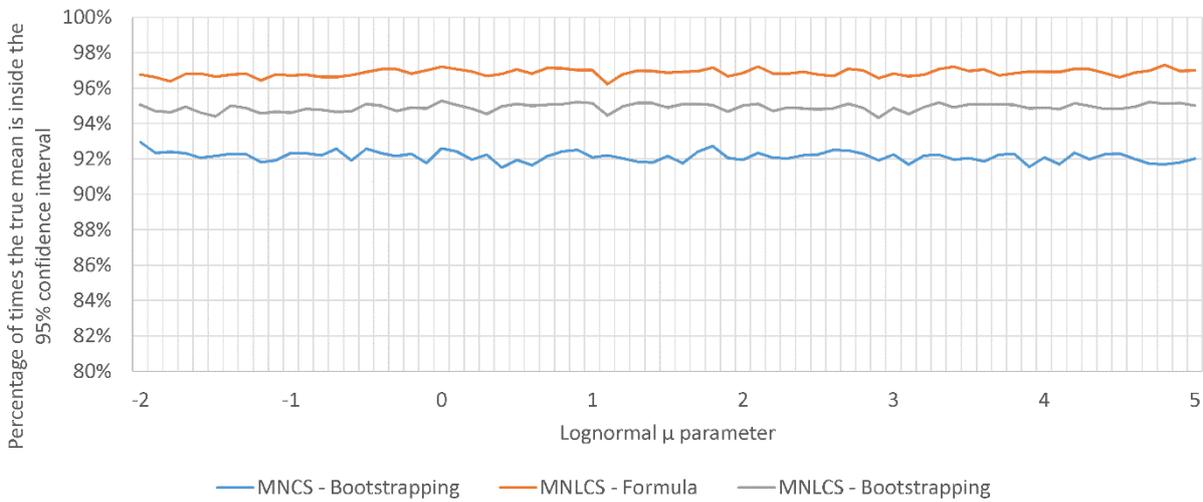

**Figure 8**. As for Figure 7 except $\sigma = 1.5$.

Changing focus to the size of the world set, bootstrapping for MNLCS becomes unsafe for smaller world sample sizes but the MNLCS formula confidence interval is safe except perhaps for fewer than 500 articles in combination with a high standard deviation (Figures 9, 10). The bootstrapping MNLCS confidence intervals are probably safe for world sample sizes of above 1000 when $\sigma = 0.75$ or for world sample sizes of above 2000 when $\sigma = 1.5$

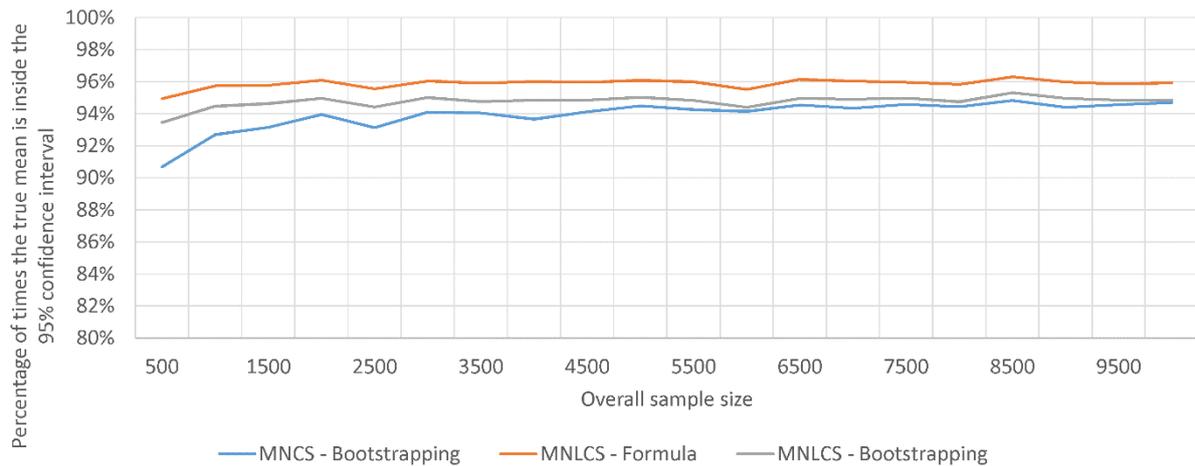

**Figure 9**. As for Figure 1 except with $\mu = 1$, $\sigma = 0.75$ fixed, the group population set at 5% and the overall sample size varying from 500 to 10000 in steps of 500.

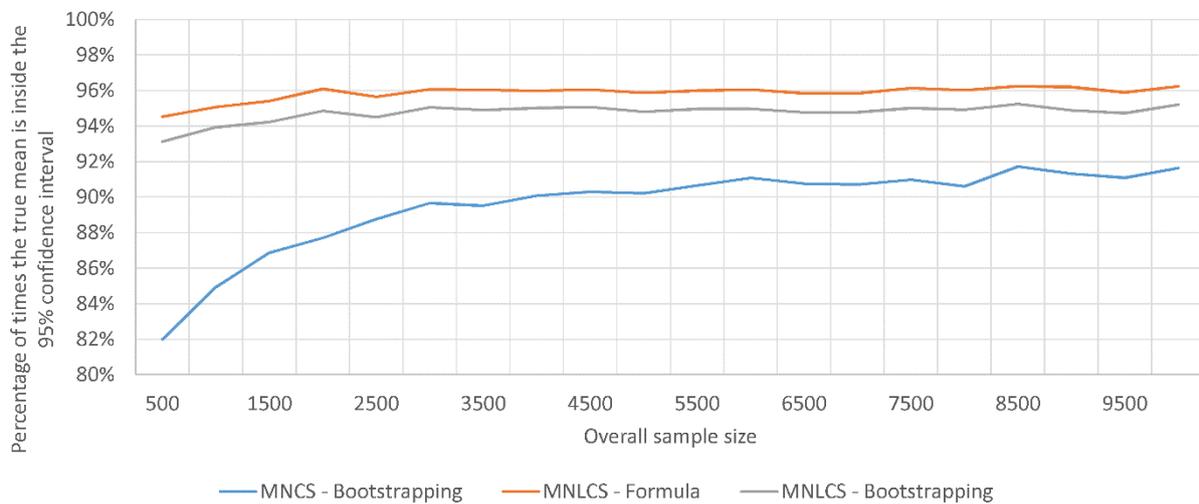

**Figure 10**. As for Figure 9 except with $\sigma = 1.5$.

# 5 Discussion

A limitation of the simulation method used is that results from a pure discretised lognormal distribution are only approximations to real citation data. Sets of citation counts may mix distributions from subfields and may include excess zeros due to the incorporation of non-scholarly documents within a database (Thelwall, 2016c). The simulation also does not deal with the complexities of articles being produced collaboratively between different research groups (Aksnes, Schneider, & Gunnarsson, 2012) or spanning different fields (Waltman et al., 2012). Moreover, the purpose of field normalised indicators is to allow different fields and years to be merged and compared with other sets whereas the experiments here simulate a single field and year. If multiple fields and/or years are merged and this increases the overall world sample sizes, then it seems likely that the safety of the bootstrap confidence intervals will increase and their accuracy would be about the same. In contrast, since the main influence on the accuracy of the formula MNLCS confidence interval is the proportion of the group set within the world set, increasing the overall sample size by combining different fields and/or years is unlikely to affect its conservatism.

In this experiment, both the world sets and the group sets were simulated from the distribution in each of the 10000 iterations. This process therefore treats each article from

12the group and from the rest of the world as a random realisation of an independent and identically distributed underlying process. In practice, this assumption is false because there are systematic influences on average citation counts, such as the authors of papers (if they produce more than one in a single year), their nationality, and field specialism and well as levels of collaboration and publishing journal prestige (Didegah & Thelwall, 2013; Hsiehchen, Espinoza, & Hsieh, 2015; Robson & Mousquès, 2016; Stegehuis, Litvak, & Waltman, 2015). In practice, therefore, sets of citation counts are not independent but are influenced by many factors.

The results suggest that the previously proposed MNLCS formula (Thelwall, 2017) is always safe, except perhaps for very small sample sizes, but that it is conservative if the group forms 10% or more of the world set. Conversely, MNLCS bootstrapping confidence intervals are unsafe for smaller sample sizes but are quite accurate even for high group proportions of 10%. Despite being unsafe for some values, MNLCS bootstrap confidence intervals are reasonably accurate most of the time. In contrast, MNCS bootstrap confidence intervals can be very inaccurate and are unsafe for a wide range of common parameter values found.

# 6 Conclusions

Based upon the above results, the follows strategy is recommended for calculating indicator confidence intervals.

- Use the MNLCS formula to calculate confidence intervals except if the group sample size is above 5% of the world sample size (i.e., Germany, UK, China, USA), *and* the world sample size is above 2000 (i.e., all except studies that focus on individual small subject categories). In the remaining cases (i.e., Germany, UK, China, USA for individual small subject categories) bootstrapping MNLCS confidence intervals should be used instead.
- Avoid calculating MNCS confidence intervals except if the discretised lognormal parameter values can be shown to fall within the ranges above for which they seem to be accurate.

A decision can also be aided by simulating with the parameters of any new dataset using the software supplied with this paper to see how accurate the two approaches are for a different parameter set.

A corollary to the above results is that the MNLCS is a robust formula for estimating average citation impact in the sense that there is always a strategy for calculating reasonably accurate confidence intervals for it. This contrasts with the situation for MNCS and similar field normalised indicators. Even if confidence intervals are not required for a given analysis, this property should give confidence in the use of the MNLCS indicator. Nevertheless, a decision to use the MNLCS implies accepting the logic that log-normalised citation counts are as good as, or better than, raw citation counts for representing the citation impact of an article.

To emphasise the importance of calculating accurate confidence intervals, the above results show that the use of an inappropriate confidence interval gives a substantial risk that a conclusion may be drawn that is not supported by the data. For example, a study with an incorrect confidence interval might find spurious statistical evidence that Spanish chemistry research had become more cited than the world average (e.g., following government restructuring), when a more accurate confidence interval would show that Spanish chemistry citation impact was within an acceptable margin of error from the world average.

Finally, if confidence intervals are used in conjunction with field normalised indicators then care should be taken to be clear about what they mean (Waltman, 2016). It is also important to state that the confidence intervals are imperfect because of the factors discussed above that can influence citation counts, as well as flaws in the article or journal classification scheme used. Moreover, if the confidence intervals give statistical evidence that one group has a higher or lower average citation rate than another group, or than the world average, then this should not be interpreted as *proving* statistically that its *research* impact or quality is higher/lower because citations reflect only one type of impact and one aspect of research quality (van Driel, Maier, & De Maeseneer, 2007).

# 8 Appendix

**Table A1**. Discretised lognormal distributions fitted to citation counts +1 for articles from 55 Scopus categories for articles published in 2011, using Scopus citation counts collected in November-December 2014 (new calculations with data reused from: Thelwall & Sud, 2016). Each subject is within one of five broad Scopus subject areas, as signalled by the bold subject names. The μ values range from -5.1 to 1.8, with a median of 1 and the σ values range from 0.9 to 2.2 with a median of 1.1. Excluding the Pharmacology, Toxicology & Pharmaceutics outlier, the smallest μ value is -0.8 and the largest σ value is 1.5.

| Subject | Articles | μ | σ |
|---|---|---|---|
| **Agricultural and Biological Sciences** (misc.) | 2329 | 0.92 | 1.16 |
| Agronomy and Crop Science | 8155 | 1.01 | 1.11 |
| Animal Science and Zoology | 8609 | 1.08 | 1.01 |
| Aquatic Science | 8901 | 1.52 | 0.92 |
| Ecology, Evolution, Behavior & Systematics | 7270 | 1.55 | 1.08 |
| Food Science | 8338 | 1.34 | 1.12 |
| Forestry | 6262 | 1.13 | 1.14 |
| Horticulture | 4334 | 1.17 | 0.99 |
| Insect Science | 6811 | 1.12 | 0.98 |
| Plant Science | 7783 | 1.32 | 1.18 |
| Soil Science | 8658 | 1.36 | 1.04 |
| **Business, Management & Accounting** (misc.) | 1668 | 0.71 | 1.00 |
| Accounting | 2977 | 1.14 | 1.12 |
| Business and International Management | 8437 | 0.50 | 1.39 |
| Management Information Systems | 2324 | 0.70 | 1.42 |
| Management of Technology & Innovation | 5256 | 1.01 | 1.24 |
| Marketing | 4513 | 1.10 | 1.14 |
| Org. Behav. & Human Resource Management | 4255 | 0.94 | 1.15 |
| Strategy and Management | 9156 | 0.85 | 1.31 |
| Tourism, Leisure & Hospitality Management | 1956 | 1.34 | 1.03 |
| Industrial Relations | 1347 | 0.02 | 1.36 |
| **Pharmacology, Toxicology & Pharmaceutics** | 534 | -5.07 | 2.21 |
| Drug Discovery | 8710 | 1.63 | 1.03 |
| Pharmaceutical Science | 1335 | 1.40 | 1.10 |
| Pharmacology | 8034 | 1.53 | 1.17 |
| Toxicology | 9169 | 1.78 | 0.94 |
| **Psychology** (miscellaneous) | 989 | 1.09 | 1.06 |
| Applied Psychology | 6208 | 1.45 | 1.02 |
| Clinical Psychology | 9085 | 1.32 | 1.14 |
| Developmental and Educational Psychology | 8805 | 1.44 | 1.06 |
| Experimental and Cognitive Psychology | 6566 | 1.74 | 0.95 |
| Neuropsychology and Physiological Psychology | 2434 | 1.75 | 0.95 |
| Social Psychology | 7509 | 1.27 | 1.06 |
| **Social Sciences** (misc.) | 5554 | 0.78 | 1.11 |
| Archeology | 2733 | 0.02 | 1.42 |
| Development | 5715 | 0.75 | 1.17 |
| Education | 7253 | 0.69 | 1.15 |
| Geography, Planning and Development | 8907 | 0.65 | 1.26 |
| Health (social science) | 8578 | 1.07 | 1.09 |
| Human Factors and Ergonomics | 1286 | 1.43 | 0.99 |
| Law | 9102 | 0.36 | 1.28 |
| Library and Information Sciences | 6206 | 0.60 | 1.33 |
| Linguistics and Language | 8544 | -0.22 | 1.47 |
| Safety Research | 1423 | 0.60 | 1.13 |
| Sociology and Political Science | 7607 | 0.53 | 1.28 |
| Transportation | 2550 | 1.32 | 1.14 |
| Anthropology | 5329 | 0.53 | 1.15 |
| Communication | 4921 | 0.57 | 1.23 |
| Cultural Studies | 9171 | -0.85 | 1.38 |
| Demography | 1802 | 0.77 | 1.21 |
| Gender Studies | 2334 | 0.63 | 1.07 |
| Life-span and Life-course Studies | 1055 | 1.17 | 1.03 |
| Political Science & International Relations | 9102 | 0.26 | 1.14 |
| Public Administration | 2908 | 0.74 | 1.10 |
| Urban Studies | 2254 | 0.65 | 1.13 |